\documentclass[prd,nofootinbib,floats,superscriptaddress,eqsecnum,tightenlines,11pt]{revtex4}
\usepackage{hyperref}
\usepackage{graphicx}
\usepackage{subfigure}
\usepackage{amsmath,amssymb,amsfonts,amsthm,latexsym,stmaryrd,bm}
\usepackage{marginnote}
\usepackage{color}
\usepackage{dcolumn}% Align table columns on decimal point
\usepackage{bm}% bold math
\newcommand{\td}{\mathrm{d}}

\begin{document}
\title{Distinguishing $f(R)$ theories from general relativity by gravitational lensing effect}%

\author{Hongguang Liu}
\email[]{liu.hongguang@cpt.univ-mrs.fr}
\affiliation{%
Department of Physics, Beijing Normal University,\\ Beijing, 100875, China
}%
\affiliation{Centre de Physique Th\'eorique (UMR 7332) , Aix Marseille Universit\'e and Universit\'e de Toulon, 13288 Marseille, France}

\author{Xin Wang}%
\author{Haida Li}%
\author{Yongge Ma}%
\email[Correspondingauthor:]{mayg@bnu.edu.cn}
\affiliation{%
Department of Physics, Beijing Normal University,\\ Beijing, 100875, China
}%

\date{\today}% It is always \today, today,

\begin{abstract}
The post-Newtonian formulation of a general class of $f(R)$ theories is set up to $3^{rd}$ order approximation. It turns out that the information of a specific form of $f(R)$ gravity is encoded in the Yukawa potential, which is contained in the  perturbative expansion of the metric components. Although the Yukawa potential is canceled in the $2^{nd}$ order expression of the effective refraction index of light, detailed analysis shows that the difference of the lensing effect between the $f(R)$ gravity and general relativity does appear at the $3^{rd}$ order when $\sqrt{f''(0)/f'(0)}$ is larger than the distance $d_0$ to the gravitational source. However, the difference between these two kinds of theories will disappear in the axially symmetric spacetime region. Therefore only in very rare case the $f(R)$ theories are distinguishable from general relativity by gravitational lensing effect at the $3^{rd}$ order post-Newtonian approximation.

\begin{description}
\item[Keywords]
$f(R)$ theory, post-Newtonion approximation, gravitational lens
\item[PACS numbers]
04.50.Kd, 04.25.Nx, 98.62.Sb
\end{description}
\end{abstract}

\pacs{Valid PACS appear here}% PACS, the Physics and Astronomy
%\keywords{}%Use showkeys class option if keyword
\maketitle

%\tableofcontents

\section{Introduction}
Recently modified gravity theories have received increasingly attention in issues related to "dark energy" \cite{riess2004type,eisenstein2005detection,astier2006supernova}, "dark matter" \cite{page2007three,zwicky1933spectral,rubin1970rotation,di2013dark}, as well as non-trivial tests on gravity beyond general relativity (GR) \cite{Will:2014bqa}. Historically, Einstein's GR is the simplest relativistic theory of gravity with correct Newtonian limit. To pursue new physics, Weyl and Eddington even began to consider modifying GR just after it was established \cite{weylei,eddington1920mathematical}. From the viewpoints of perturbutive quantum gravity, GR is non-renormalizable \cite{utiyama1962renormalization,stelle1977renormalization,buchbinder1992effective,vilkovisky1992effective}, while higher order gravity theories might alleviate the problem. From the phenomenological viewpoints, there are many ways to modify GR, and some empirical approaches seem to have promising prospect, such as Dvali-Gabadadze-Porrati gravity \cite{dvali20004d}, tensor-vector-scalar theory \cite{fujii2003scalar} and Einstein-Aether theory \cite{jacobson2001gravity}. Among such extended theories, particular attention has been devoted to the so-called $f(R)$-gravity. This kind of theories is based on a generalization of the Einstein Hilbert Lagrangian to nonlinear functions $f(R)$ of the Ricci scalar \cite{sotiriou2010f}. $f(R)$-gravity covers a lot of characteristics of higher order gravity and is convenient to be operated. Hence, $f(R)$ theories provide an ideal tool to study the possible extension of GR. $f(R)$ theories of gravity can also be non-perturbatively quantized by loop quantum gravity approach \cite{Zhang:2011vi,Zhang:2011qq}.
%cleared

To confront $f(R)$-gravity with observations in Solar System, one can get constraints on the theories from different measurements, such as the E\"{o}tWash experiment \cite{kapner2007tests}, the geodesic precession of gyroscopes measured by Gravity Probe B \cite{Everitt:2011hp} and the precession of the binary pulsars PSR J0737-3039 \cite{Breton:2008xy}. At cosmological scales one would expect  to employ $f(R)$ theories to account for the problems of "dark energy" \cite{nojiri2003modified,capozziello2003curvature,Huang:2009vw} and "dark matter" \cite{sobouti2007mathsf,capozziello2009testing,capozziello2012dark} needed in GR. If $f(R)$ gravity could account for dark matter, besides matching the rotation curves of galaxy clusters, it should also match the measurements on gravitational lensing effect \cite{schneider1996detection}. However, it is shown in \cite{Lubini11} that, at $2^{nd}$ order post-Newtonian approximation, a rather general class of $f(R)$ theories is indistinguishable from GR in gravitational lensing effect.
 Nevertheless, we will show in this paper that a class of $f(R)$ theories is indeed distinguishable from GR in gravitational lensing effect at $3^{rd}$ order post-Newtonian approximation. However, the possibility to account for the dark matter problem with $f(R)$ theory in lensing effect is highly suppressed due to this tiny $3^{rd}$ order difference.
%cleared

This paper is organized as follows. In section \ref{ch:field_equations_fr}, we briefly review the field equations of metric $f(R)$-gravity. In section \ref{ch:post_newtonian_expansion} the post-Newtonian approximation of a class of $f(R)$ theories is formulated to the desired order. In section \ref{ch:gravitational_lensing} we introduce the gravitational lensing effect in metric theories of gravity and show how the $f(R)$ gravity can be distinguishable from GR at $3^{rd}$-order post-Newtonian approximation. The difference of the lensing reflection indexes is discussed in an example. Finally, conclusions and remarks are given in Sec. \ref{ch:summary}. Throughout the paper, the metric tensor $g_{\mu\nu}$ takes the signature $(-,+,+,+)$.
%cleared

\section{Field Equations of $f(R)$ theory}\label{ch:field_equations_fr}

In metric $f(R)$ theories of gravity, the action of gravity coupled to matter fields is given by
\begin{equation}\label{eq:action}
S=\frac{1}{2\chi}{\int}f(R)\sqrt{-g} \td ^{4}x+S_{M} \;,
\end{equation}
where $g$ is the determinant of the metric tensor $g_{{\mu}{\nu}}$, $\chi=8{\pi}G/c^{4}$ with $G$ and $c$ being the Newtonian gravitational constant and the vacuum speed of light respectively, $R=g^{{\mu}{\nu}}R_{{\mu}{\nu}}$ is the Ricci scalar, $f(R)$ is a nonlinear function and $S_{M}$ is the standard matter action. The variation of action (\ref{eq:action}) with respect to the metric $g_{\mu{\nu}}$ yields the Euler-Lagrange
equations
\begin{equation}\label{eq:field_equations}
\begin{split}
&f'(R)R_{\mu \nu}-\frac{1}{2}f(R)g_{\mu \nu}-\nabla_{\mu}\nabla_{\nu}f'(R)+g_{\mu\nu}\square_gf'(R)={\chi}T_{\mu\nu} \;,
\end{split}
\end{equation}
where $\nabla_{\mu}$ is the covariant derivative for $g_{{\mu}{\nu}}$, $\square_g:=\nabla^{\mu}\nabla_{\mu}$, and $T_{{\mu}{\nu}}=(-2c/\sqrt{-g})(\delta{S}_{M}/{\delta}{g}^{{\mu}{\nu}})$ is the energy momentum tensor of matter.
Taking the trace of Eq. (\ref{eq:field_equations}) we can get
\begin{equation}\label{eq:trace}
3\square_gf'(R)+f'(R)R-2f(R)={\chi}T \;,
\end{equation}
where $T$ is the trace of $T_{{\mu}{\nu}}$. Using Eq. (\ref{eq:trace}), we can rewrite Eq. (\ref{eq:field_equations}) as
\begin{equation}\label{eq:field_equations_new}
\begin{split}
&R_{\mu \nu}=\frac{1}{f'(R)}(\frac{1}{3}g_{\mu\nu}f'(R)R-\frac{1}{6}f(R)g_{\mu \nu}+\nabla_{\mu}\nabla_{\nu}f'(R)+{\chi}(T_{\mu\nu}-\frac{1}{3}g_{\mu\nu}T)) \;.
\end{split}
\end{equation}
%cleared

\section{Post-Newtonian Expansion}\label{ch:post_newtonian_expansion}

The matter constituents in the universe are usually well approximated by a perfect fluid with mass density $\rho$ and pressure $p$ \cite{weinberg1972gravitation}. Hence we assume that the Newtonian potential $U$ of the mass distribution, the typical velocities $v$ and the pressure of the fluid obey such approximation respectively. In the post-Newtonian approximation, we can further expand the dynamical variables in the field equations perturbatively in powers of $1/c$, since we have the the following order relation \cite{will1993theory,clifton2008parametrized,stabile2010post}
\begin{equation}\label{eq:expansion}
\frac{U }{{{c}^{2}}}\sim \frac{{{v}^{2}}}{{{c}^{2}}}\sim \frac{p}{\rho {{c}^{2}}}\sim \Pi \sim O({2}) \;,
\end{equation}
where $\Pi$ is the ratio of the energy density to the rest-mass density.
%cleared

We consider the case that the gravitational field is weak and assume that in absence of a gravitational field the background space-time is flat \cite{schneider1996detection}. We also assume that $f(0) = 0$, which neglects the contribution of a possible cosmological constant and excludes some form of $f(R)$ theories, e.g, $f(R)=1/R$. Note that actually the contribution of a possible cosmological constant can be equivalently substituted by the corresponding contribution of a energy-momentum tensor. Moreover, the $f(R)$ form which is unable to get weak field solution is useless here. In a weak field regime the metric tensor can be expanded about the Minkowski metric $\eta_{\mu\nu}$ in its Lorentzian coordinate system as
\begin{equation}
g_{{\mu}{\nu}}=\eta_{\mu\nu}+h_{\mu\nu} \;,
\end{equation}
where $|h_{\mu\nu}|\ll1$. Up to $3^{rd}$ order the components of the metric tensor can be written as \cite{naf20101,Lubini11,will1993theory}:
\begin{equation}\label{eq:metric}
\begin{split}
{{g}_{00}}=&-1{{+}^{(2)}}{{h}_{00}}+O(4) \;,\\
{{g}_{0i}}=&^{(3)}{{h}_{0i}}+O(5) \;,\\
{{g}_{ij}}=&{{\delta }_{ij}}{{+}^{(2)}}{{h}_{ij}}+O(4) \;,\\
\end{split}
\end{equation}
where the left upper index (n) means the order $O(n)$. Using Eq. (\ref{eq:metric}) we can get the components of the Ricci tensor as
\begin{equation}\label{eq:ricci_tensor}
\begin{split}
{{R}_{00}}=&-\frac{1}{2} {{\nabla}^{2}}{^{(2)}}{h_{00}}+O(4) \;,\\
{{R}_{0i}}=&\frac{1}{2}(-{{\nabla}^{2}}{^{(3)}}{h_{0i}}-\frac{1}{c}{^{(2)}h_{jj,0i}}+{^{(3)}h_{j0,ij}}+\frac{1}{c}{^{(2)}h_{ij,0j}})+O(5) \;,\\
{{R}_{ij}}=&\frac{1}{2}(-{{\nabla}^{2}}{^{(2)}}{h_{ij}}+{^{(2)}h_{00,ij}}-{^{(2)}h_{kk,ij}}+{^{(2)}h_{ik,kj}}+{^{(2)}h_{kj,ki}})+O(4) \;.\\
\end{split}
\end{equation}
Assuming $f(R)$ to be analytic at $R = 0$, to the $2^{nd}$ order the Ricci scalar and thus $f(R)$ and $f'(R)$ read respectively as
\begin{equation}\label{eq:fr_expansion}
\begin{split}
R &=^{(2)}R+O(4) \;,\\
f(R) &=f'{{(0)}^{(2)}}R+O(4) \;,\\
f'(R) &=f'(0)+f''{{(0)}^{(2)}}R+O(4) \;.\\
\end{split}
\end{equation}
To the leading order the components of the energy-momentum tensor of matter fields read
\begin{equation}
\begin{split}
&{\chi{T}^{00}}=\chi{^{(-2)}}{T^{00}}+O(4) \;,\\
&{\chi{T}^{0i}}=\chi{^{(-1)}}{T^{0i}}+O(4) \;,\\
&{\chi{T}^{ij}}=O(4) \;.\\
\end{split}
\end{equation}
Note that if $f''(0)=0$, Eq. (\ref{eq:fr_expansion}) implies that $f'(R)$ is constant. Then Eq. (\ref{eq:trace}) yields at the $2^{nd}$ order $^{(2)}R=-\chi^{(2)}T_{00}/f'(0)$, which is consistent with the equation of GR at the same order. Thus in this approach, GR is nothing else but the first term of the Taylor expansion of a more general $f(R)$ theory. As one can see from the action (\ref{eq:action}), we ask $f(R)$ to carry the same dimension as $R$'s. Thus both $f'(R)$ and the term $f''(0) ^{(2)}R$ in Eq. (\ref{eq:fr_expansion}) are dimensionless. Since the term $f''(0) ^{(2)}R$ is required to be of order $O(2)$, the expansion of Eq. (\ref{eq:fr_expansion}) will break down if $^{(2)}R {\geq}f'(0)/f''(0)$ \cite{Lubini11}.
%cleared

To derive neat equations in the post-Newtonian approximation, we impose the gauge conditions \cite{naf20101,Lubini11}
\begin{align}
& g_{ij,j}-\frac{1}{2}(g_{jj}-g_{00})_{,i}-\frac{f'(R)_{,i}}{f'(R)}=O(4) \;,\\
& g_{0j,j}-\frac{1}{2c}g_{jj,0}-\frac{f'(R)_{,0}}{cf'(R)}=O(5) \;.
\end{align} 
%cleared

With the gauge conditions, we get from Eqs. (\ref{eq:trace}) and (\ref{eq:field_equations_new})
\begin{align}\label{eq:dr}
& {{\nabla }^{2(2)}}R=-\frac{1}{3f''(0)}{{\chi }^{(-2)}}{{T}^{00}}+{{{\frac{f'(0)}{3f''(0)}}}{^{(2)}}R} \;,\\
\label{eq:dh00}
& {{\nabla }^{2(2)}}{{h}_{00}}=-{\frac{4 \chi }{3f'(0)}}{^{(-2)}{T}^{00}}+\frac{1}{3}{^{(2)}}R \;,\\
\label{eq:dhij}
& {{\nabla }^{2(2)}}{{h}_{ij}}=-({\frac{2\chi }{3f'(0)}}{^{(-2)}{{T}^{00}}}+\frac{1}{3}{^{(2)}}R){{\delta }_{ij}} \;,\\
\label{eq:dh0i}
& \nabla^2{^{(3)}{h_{0i}}}=\frac{2\chi}{f'(0)}{^{(-1)}T^{0i}}-\frac{1}{2c}{^{(2)}h_{00,0i}} \;.
\end{align}
For the sake of physics and simplicity, we consider the case of $f'(0)>0$ and $f''(0)>0$ and define $\alpha^2:=f'(0)/(3f''(0))$. Note that in this case the constant $1/f'(0)$ can be absorbed into the gravitational constant $G$ if necessary. Then from Eq. (\ref{eq:dr}) we can get \cite{clifton2008parametrized,naf20101}
\begin{equation}\label{eq:r}
{^{(2)}R}=\frac{2\alpha^2}{c^2}V(\bm{x},t) \;,
\end{equation}
with the Yukawa potential
\begin{equation}\label{eq:V}
V(\bm{x},t):=\frac{G}{c^2}\int{\frac{{^{(-2)}T^{00}(\bm{x}',t)}\mathrm{e}^{-\alpha \vert \bm{x}-\bm{x}' \vert}}{\vert \bm{x}-\bm{x}' \vert}} \td ^3 x' \;.
\end{equation}
Note that the information of a specific form of $f(R)$ gravity is encoded in the parameter $\alpha$ in the potential $V$, and we only consider the solution with $\alpha>0$. It should be noticed that, for the other solution with $\alpha<0$, the potential $V$ would tend to be divergent at infinity.

 It is easy to show by using Eq. (\ref{eq:dr}) that Eq. (\ref{eq:dh00}) can be written as
\begin{equation}\label{eq:dh00_new}
{{\nabla }^{2(2)}}{{h}_{00}}=-{\frac{\chi }{f'(0)}} {^{(-2)}{{T}^{00}}}+\frac{f''(0)}{f'(0)}{{\nabla }^{2(2)}}R \;.
\end{equation}
Thus we get
\begin{equation}\label{eq:dh00_1}
{{\nabla }^{2}}(^{(2)}{{h}_{00}}-\frac{1}{3\alpha{^2}}{^{(2)}}R)=-{\frac{\chi }{f'(0)}} {^{(-2)}{{T}^{00}}} \;.
\end{equation}
Using Eq. (\ref{eq:r}), the solution of Eq. (\ref{eq:dh00_1}) can be given by
\begin{equation}\label{eq:h002}
{^{(2)}h_{00}(\bm{x},t)}=\frac{2}{c^2}(U(\bm{x},t)+\frac{1}{3}V(\bm{x},t)) \;,
\end{equation}
where the Newtonian potential $U$ reads
\begin{equation}\label{eq:U}
U(\bm{x},t):=\frac{G}{c^2}\int{\frac{{^{(-2)}T^{00}(\bm{x}',t)}}{\vert \bm{x}-\bm{x}' \vert}} \td ^3 x' \;.
\end{equation}
It is obvious that the Newtonian potential $U$ remains unchanged for different forms of $f(R)$ gravity. Similarly, the solution of Eq. (\ref{eq:dhij}) reads
\begin{equation}\label{eq:hij2}
{^{(2)}h_{ij}(\bm{x},t)}=\frac{2\delta_{ij}}{c^2}(U(\bm{x},t)-\frac{1}{3}V(\bm{x},t)) \;.
\end{equation}
From Eqs. (\ref{eq:dh00}) and (\ref{eq:dhij}), we have
\begin{equation}
{\nabla ^2 }(^{(2)}{h}_{00}-2^{(2)}{{h}_{ii}})=-\frac{4 \chi }{3f'(0)} {^{(-2)}{T}^{00}}+\frac{1}{3}{^{(2)}}R+2({\frac{2\chi }{3f'(0)}} {^{(-2)}{T}^{00}}+\frac{1}{3}{^{(2)}}R)={^{(2)}R} \;.
\end{equation}
Using Eqs. (\ref{eq:h002}) and (\ref{eq:hij2}), it is easy to get
\begin{equation}
{\nabla ^2 }(^{(2)}{h}_{00}-2^{(2)}{{h}_{ii}})=\frac{2}{c^2}{\nabla ^2 }(V-U) \;.
\end{equation}
Thus from Eq. (\ref{eq:r}) we have
\begin{equation}
V=\frac{1}{{\alpha}^2}{\nabla ^2 }(V-U) \;.
\end{equation}
Then Eq.(\ref{eq:h002}) can be written as
\begin{equation}\label{eq:h002_new}
{^{(2)}h_{00}(\bm{x},t)}=\frac{1}{c^2}{\nabla ^2 }(-\psi+\frac{2}{{3\alpha}^2}(V-U)) \;,
\end{equation}
where the potential $\psi$ is defined as
\begin{equation}
\begin{split}
\psi(\bm{x},t):=&-\frac{G}{c^2}\int{{^{(-2)}T^{00}(\bm{x}',t)}\vert \bm{x}-\bm{x}' \vert} \td ^3x' \;,
\end{split}
\end{equation}
such that ${\nabla ^2 }\psi=-2U$. Hence Eq. (\ref{eq:dh0i}) can be written as
\begin{equation}\label{eq:dh0i_new}
\nabla^2({^{(3)}h_{0i}}+\frac{1}{2c^3}(-\psi+\frac{2(V-U)}{3\alpha{^2}})_{,0i})=\frac{2{\chi}}{f'(0)}{^{(-1)}T^{0i}} \;.
\end{equation}
The solution of Eq. (\ref{eq:dh0i_new}) reads
\begin{equation}\label{eq:h0i}
{^{(3)}h_{0i}(\bm{x},t)}=\frac{1}{c^3}(-4 Y_i(\bm{x},t)+\frac{1}{2}\psi(\bm{x},t)_{,0i}+Z(\bm{x},t)_{,0i}) \;,
\end{equation}
where
\begin{equation}\label{eq:y_z}
\begin{split}
Y_i(\bm{x},t):=&\frac{G}{c}\int{\frac{{^{(-1)}T^{0i}}(\bm{x}',t)}{\vert \bm{x}-\bm{x}' \vert}} \td ^3x' \;,\\
Z(\bm{x},t):=&\frac{1}{3\alpha{^2}}(U(\bm{x},t)-V(\bm{x},t)) \;.
\end{split}
\end{equation}

So up to $3^{rd}$ order post-Newtonian approximation, the final form of the metric components reads
\begin{equation}\label{eq:fr_metric}
\begin{split}
{g_{00}}&=-1+\frac{2}{c^2}U+\frac{2}{3c^2}V \;,\\
{g_{0i}}&=\frac{1}{c^3}(-4 Y_i+\frac{1}{2}\psi_{,0i}+Z_{,0i}) \;,\\
{g_{ij}}&=(1+\frac{2}{c^2}U-\frac{2}{3c^2}V){\delta_{ij}} \;.\\
\end{split}
\end{equation}

In contrast, the metric components to the same order approximation in GR reads \cite{will1993theory}
\begin{equation}\label{eq:gr_metric}
\begin{split}
{g_{00}}&=-1+\frac{2}{c^2}U \;,\\
{g_{0i}}&=\frac{1}{c^3}(-4 Y_i+\frac{1}{2}\psi_{,0i}) \;,\\
{g_{ij}}&=(1+\frac{2}{c^2}U){\delta_{ij}} \;.\\
\end{split}
\end{equation}
Hence the difference between the $f(R)$ gravity and GR comes from the Yukawa-like potential $V$ and $Z_{,0i}$. In the limit $f''(0) \to 0$, we get $\alpha \to \infty$ and $V \to 0$. Then the solution (\ref{eq:fr_metric}) of $f(R)$ gravity goes back to (\ref{eq:gr_metric}) of GR. On the other hand, it is straightforward to see that, in the limit $f''(0) \to \infty$, we have $\alpha \to 0$ and hence get the most obvious departure of $f(R)$ gravity from GR.
%cleared

\section{Gravitational Lensing}\label{ch:gravitational_lensing}

A gravitational lens refers to a distribution of plates (such as a cluster of galaxies) between a distant source (a background galaxy) and an observer, that is capable of bending the light from the source, as it travels towards the observer.
The lensing effect can magnify and distort the image of the background source \cite{schneider1992gravitational}. According to Fermat's principle, the world line of a light should extremize its arrival time $T$ with respect to an observer under the variation of $\gamma$. In metric theories of gravity, this principle implies that the world line of a light coincides with a null geodesic in the spacetime. In the Lorentzian coordinate system of the flat background spacetime, let  $dl^2 = \delta_{ij} \td  x^i d x^j$ be the spatial Euclidean line element. Up to a constant, the travel time of light on a null geodesic $\gamma$ is given by
\begin{equation}\label{eq:light_path}
T=\int_{\gamma } \td t\ \text{=}\int_{\gamma }\frac{\td t}{\td l}\td l\ \text{=}\ \frac{1}{c}\int_{\gamma }n\ \td l \;,
\end{equation}
where we defined the effective refraction index of light as
\begin{equation}\label{eq:n_def}
n:=c\frac{dt}{\td l} \;.
\end{equation}
Then Eq.(\ref{eq:light_path}) takes the form similar to that of the propagation of a light through a medium in Newtonian space and time.
%cleared

\subsection{$2^{nd}$ Order Expansion}
At the $2^{nd}$ order post-Newtonian approximation, the only nonzero perturbative metric components are $h_{00}$ and $h_{ij}$ in Eqs. (\ref{eq:h002}) and (\ref{eq:hij2}). For a null geodesic, we have \cite{schneider2006gravitational, Lubini11, blanchet2001general}
\begin{equation}\label{eq:ds_2}
\td{{s}^{2}}=-(1-{}^{(2)}{{h}_{00}}){{c}^{2}}\text{d}{{t}^{2}}+({{\delta }_{ij}}+{}^{(2)}{{h}_{ij}})\td {{x}^{i}} \td{{x}^{j}}=0 \;.
\end{equation}
Hence we can get
\begin{equation}
{{c}^{2}}\td{{t}^{2}}=\frac{{{\delta }_{ij}}+^{(2)}{{h}_{ij}}}{1{{-}^{(2)}}{{h}_{00}}} \td {{x}^{i}} \td {{x}^{j}}=\frac{1+^{(2)}h}{1-^{(2)}{{h}_{00}}}\td{{l}^{2}} \;,
\end{equation}
where $h$ is defined such that Eq. (\ref{eq:hij2}) can be written as $^{(2)}{h}_{ij} = ^{(2)}h{\delta}_{ij}$. Using Eq. (\ref{eq:n_def}), we can obtain the effective index of refraction as
\begin{equation}
n=c\frac{\td t}{\td l}=\sqrt{\frac{1+(2)h}{1-(2){{h}_{00}}}} \;,
\end{equation}
which at the $2^{nd}$ order reads
\begin{equation}
n=1+\frac{1}{2}{{(}^{(2)}}{{h}_{00}}{{+}^{(2)}}h) \;.
\end{equation}
By Eqs. (\ref{eq:h002}) and (\ref{eq:hij2}), it equals to
\begin{equation}
n=1+\frac{2}{c^2}U(x,t) \;,
\end{equation}
which illustrates that at $2^{nd}$ order, the effective refraction index $n$ is only ¥determined by the Newtonian potential $U$. As shown in Eqs. (\ref{eq:fr_metric}) and (\ref{eq:gr_metric}), the difference between $f(R)$ gravity and GR comes from the potential $V$ rather than $U$. Hence one can not distinguish $f(R)$ theories from GR at $2^{nd}$ order approximation by the gravitational lensing effect \cite{Lubini11}.
%cleared

\subsection{$3^{rd}$ Order Expansion}
We now consider the $3^{rd}$ order post-Newtonian approximation which is needed in dealing with light rays in spacetime \cite{will1993theory}.
At the $3^{rd}$ order expansion, the line element of the metric can be written as
\begin{equation}
\begin{split}
\td{{s}^{2}}=&-(1-{}^{(2)}{{h}_{00}}){{c}^{2}}\text{d}{{t}^{2}}+({{\delta }_{ij}}+^{(2)}{{h}_{ij}}) \td{{x}^{i}} \td {{x}^{j}}+2^{(3)}{{h}_{0i}}c dtd{{x}^{i}} \;.\\
\end{split}
\end{equation}
For a null geodesic, by using Eq. (\ref{eq:n_def}) we have
\begin{equation}
\begin{split}
0=&-(1-{}^{(2)}{{h}_{00}}){n}^2+({{\delta }_{ij}}+^{(2)}{{h}_{ij}})\frac{d{{x}^{i}} \td {{x}^{j}}}{{\td l}^2}+2^{(3)}{{h}_{0i}}\frac{d{{x}^{i}}}{\td l}n \;.\\
\end{split}
\end{equation}
Thus we obtain
\begin{equation}\label{eq:n_3}
n=-\frac{{^{(3)}h_{0i}}}{-1+{^{(2)}h_{00}}}\frac{\td x^i}{\td l}
\pm{F({^{(2)}h_{00}},{^{(2)}h},{^{(3)}h_{0i}})} \;,
\end{equation}
where
\begin{equation}\label{eq:F}
\begin{split}
F:=&\sqrt{\frac{({^{(3)}h_{0i}}\td x^i)^2}{(1-{^{(2)}h_{00}})^2{\td l}^2}
-\frac{(1+{^{(2)}h})({\td x^i}^2)}{(-1+{^{(2)}h_{00}}){\td l}^2}}=\sqrt{\frac{({^{(3)}h_{0i}}\td x^i)^2}{(1-{^{(2)}h_{00}})^2{\td l}^2}
+\frac{1+{^{(2)}h}}{1-{^{(2)}h_{00}}}} \;.\\
\end{split}
\end{equation}
At $3^{rd}$ order approximation, Eq. (\ref{eq:F}) can be expressed as
\begin{equation}
\begin{split}
F({^{(2)}h_{00}},{^{(2)}h},{^{(3)}h_{0i}})=&F|_{0,0,0}+\frac{\partial F}{\partial {^{(2)}h_{00}}}|_{(0,0,0)}{^{(2)}h_{00}}+\frac{\partial F}{\partial {^{(2)}h}}|_{(0,0,0)}{^{(2)}h}+\frac{\partial F}{\partial {^{(3)}h_{0i}}}|_{(0,0,0)}{^{(3)}h_{0i}}\\
&+{^{(4)}E({^{(2)}h_{00}},{^{(2)}h})}+O(5)\\
=&1+\frac{1}{2}
{^{(2)}h_{00}}+\frac{1}{2}{^{(2)}h}+O(4) \;,
\end{split}
\end{equation}
where ${^{(4)}E({^{(2)}h_{00}},{^{(2)}h})}$ represents the expansion terms at fourth order.
It is obvious that the "$-$" sign in front of the function $F$ in  Eq.(\ref{eq:n_3}) should be neglected, since otherwise the refraction index $n$ would become negative. Hence  Eq. (\ref{eq:n_3}) becomes
\begin{equation}\label{eq:n_3_new}
\begin{split}
n=&-\frac{{^{(3)}h_{0i}}\frac{\td x^i}{\td l}}{-1+{^{(2)}h_{00}}}+(1+\frac{1}{2}
{^{(2)}h_{00}}+\frac{1}{2}{^{(2)}h})={^{(3)}h_{0i}}\frac{\td x^i}{\td t}\frac{n}{c}+(1+\frac{1}{2}
{^{(2)}h_{00}}+\frac{1}{2}{^{(2)}h}) \;.
\end{split}
\end{equation}
It is easy to see that Eq. (\ref{eq:n_3_new}) can be solved as
\begin{equation}\label{eq:3_lengsing_index}
\begin{split}
n&=1+\frac{1}{2}
{^{(2)}h_{00}}+\frac{1}{2}{^{(2)}h}+\frac{1}{c}{^{(3)}h_{0i}}\frac{\td x^i}{dt}=n_2+\frac{1}{c}{^{(3)}h_{0i}}{u}^{i} \;,
\end{split}
\end{equation}
where ${u}^{i}=\td x^i/dt$ is the components of the coordinate speed of light, and $n_2:=1+2U(x,t)/{c^2}$ is the refraction index at $2^{nd}$ order.
 Therefore, at $3^{rd}$ order post-Newtonian approximation, the effective refraction index of light is obviously dependent on the $3^{rd}$ order metric components $h_{0i}$. From Eqs. (\ref{eq:y_z}), (\ref{eq:fr_metric}) and (\ref{eq:gr_metric}) one can see that, in contrast to the case of GR, in $f(R)$ gravity $h_{0i}$ is effected also by the Yukawa potential $V$. Hence, $f(R)$ theories are in principle distinguishable from GR by gravitational lensing effect at $3^{rd}$ order post-Newtonian approximation.
%cleared

\subsection{Differences: An example}
Although the difference of lensing effect between $f(R)$ gravity and GR is encoded in the $3^{rd}$ order terms, it is still unclear whether the difference can actually be detected at this order and in which case the departure become most obvious. To answer these questions, we first recall from Eq.(\ref{eq:3_lengsing_index}) that the difference at $3^{rd}$ order effect is contained in the difference of the metric components $h_{0j}$ between $f(R)$ gravity and GR, which reads
\begin{equation}\label{eq:n_diff}
\begin{split}
\Delta n:=\frac{u^j}{c}\Delta n_j=\frac{u^j}{c^4}Z_{,0j}u^j:=&\frac{1}{3\alpha{^2} c^4}(U(\bm{x},t)-V(\bm{x},t))_{,0j} \;.\\
\end{split}
\end{equation}
A straight-forward calculation leads to
\begin{equation}\label{eq:n_diff_com}
\begin{split}
\Delta n_{j}=&\frac{ G}{c^5}\int{\frac{{^{(-2)}T^{00}(\bm{x}',t)} v'_j }{\vert \bm{x}-\bm{x}' \vert} } g(\alpha \vert \bm{x}-\bm{x}' \vert)  \td^3 x'\\
&-\frac{G}{c{^5}}\int{ \frac{{^{(-2)}T^{00}(\bm{x}',t)} v' \cdot (\bm{x}-\bm{x}')(\bm{x}-\bm{x}')_j}{\vert \bm{x}-\bm{x}' \vert ^3}h(\alpha \vert \bm{x}-\bm{x}' \vert)  } \td^3 x' \;,\\
\end{split}
\end{equation}
where
\begin{equation}\label{eq:gh}
  \begin{split}
    g(\alpha \vert \bm{x}-\bm{x}' \vert)&:=\frac{1}{3 \alpha^2 \vert \bm{x}-\bm{x}' \vert^2}\bigg(1-\mathrm{e}^{-\alpha \vert \bm{x}-\bm{x}' \vert} (1 + \alpha \vert \bm{x}-\bm{x}' \vert) \bigg) \;,\\
    h(\alpha \vert \bm{x}-\bm{x}' \vert)&:=\frac{1}{\alpha^2 \vert \bm{x}-\bm{x}' \vert^2} \left(1-\mathrm{e}^{-\alpha \vert \bm{x}-\bm{x}'\vert} \bigg(1+\alpha \vert \bm{x}-\bm{x}' \vert +\frac{1}{3}\alpha^2 \vert \bm{x}-\bm{x}' \vert^2 \bigg) \right) \;
  \end{split}
\end{equation}
are two monotone decreasing functions. Hence the lensing refraction indexes of $f(R)$ gravity and GR will take the biggest departure in the limit of $\alpha \to 0$, which reads
\begin{equation}\label{eq:z_diff_limit}
\begin{split}
\Delta n_j \sim &\frac{G}{6c^5}\int{\frac{{^{(-2)}T^{00}(\bm{x}',t)} v'_j }{\vert \bm{x}-\bm{x}' \vert} }  \td^3 x'-\frac{G}{6c^5}\int{ \frac{{^{(-2)}T^{00}(\bm{x}',t)} v' \cdot (\bm{x}-\bm{x}')(\bm{x}-\bm{x}')_j}{\vert \bm{x}-\bm{x}' \vert ^3} } \td^3 x'=\frac{1}{6c^3} \psi_{,0i} \;.
\end{split}
\end{equation}
One may noticed that this is nothing else but the potential $\psi_{,0i}$ appearing in the $3^{rd}$ order post Newtonian approximation of GR. Thus, in the case of the most departure, the difference of the lensing refraction indexes is at the same $3^{rd}$ of GR.

By noticing that the functions $g$ and $h$ satisfy the relation
\begin{equation}
  \frac{h(\alpha d)}{d^3}=-\frac{1}{d}\frac{\partial [g(\alpha d)/d]}{\partial d},
\end{equation}
where $d=|\bm{x}- \bm{x}'|$, the expression (4.16) can be further simplified. In terms of cylindrical coordinates $\{r,\theta,z\}$, Eq. (4.16) can be written as
\begin{equation}\label{eq:n_diff_axis}
\begin{split}
\Delta n_{j}=&\frac{ G}{c^5}\int{}r' \td r' \td z' \int{\frac{{^{(-2)}T^{00}(\bm{x}',t)} v'_j }{\vert \bm{x}-\bm{x}' \vert} } g(\alpha \vert \bm{x}-\bm{x}' \vert)  \td \theta'\\
&+\frac{G}{c{^5}}\int{} r' \td r' \td z' \int{} \frac{{^{(-2)}T^{00}(\bm{x}',t)} v' \cdot (\bm{x}-\bm{x}')(\bm{x}-\bm{x}')_j}{r r' sin(\theta')} \frac{\td}{\td \theta'} \left( \frac{g(\alpha \vert \bm{x}-\bm{x}' \vert)}{\vert \bm{x}-\bm{x}' \vert} \right) \td \theta' \\
=&\int{} r' \td r' \td z' \int{} \left( {^{(-2)}T^{00}(\bm{x}',t)} v'_j - \frac{\td}{\td \theta'} \bigg(\frac{{^{(-2)}T^{00}(\bm{x}',t)} v' \cdot (\bm{x}-\bm{x}')(\bm{x}-\bm{x}')_j}{r r' sin(\theta')} \bigg) \right) \frac{g(\alpha \vert \bm{x}-\bm{x}' \vert)}{\vert \bm{x}-\bm{x}' \vert} \td \theta'\; ,
\end{split}
\end{equation}
where we used the identity $\td_{\theta'} |\bm{x}- \bm{x}'|=\td_{\theta'} \sqrt{r^2+r'^2-2 r r' \cos(\theta') + (z-z')^2}=r r' \sin(\theta')/|\bm{x}- \bm{x}'|$ and $\theta=0$.

In an axially symmetric spacetime, it is reasonable to consider the case that the velocities $v$ of the gravitational sources are all tangent to the $r-\theta$ plane. Then from Eq.(\ref{eq:n_diff_axis}) one gets
\begin{equation}
  \begin{split}
    \Delta n_r=&\int{} r' \td r' \td z' \int{} \Bigg( {-^{(-2)}T^{00}(\bm{x}',t)} v' \sin(\theta')\\
    & + \frac{\td}{\td \theta'} \bigg(\frac{{^{(-2)}T^{00}(\bm{x}',t)} (v' \sin(\theta')r)(r - r' cos(\theta'))}{r r' \sin(\theta)} \bigg) \Bigg) \frac{g(\alpha \vert \bm{x}-\bm{x}' \vert)}{\vert \bm{x}-\bm{x}' \vert} \td \theta'\\
    =&0 \; ,\\
  \Delta n_\theta=&\int{} r' \td r' \td z' \int{} \Bigg( {^{(-2)}T^{00}(\bm{x}',t)} v' \cos(\theta') \\
  &+ \frac{\td}{\td \theta'} \bigg(\frac{{^{(-2)}T^{00}(\bm{x}',t)} (v' \sin(\theta')r)( - r' \sin(\theta'))}{r r' \sin(\theta')} \bigg) \Bigg) \frac{g(\alpha \vert \bm{x}-\bm{x}' \vert)}{\vert \bm{x}-\bm{x}' \vert} \td \theta' \\
  =&0 \; .\\
\end{split}
\end{equation}
Therefore, in an axially symmetric spacetime region, which coincides in most cases with those of galaxies and compact objects, one can not distinguish $f(R)$ theories from GR by the lensing correction to the $3^{rd}$ order term. This result suggests that there is few opportunities to distinguish $f(R)$ theory from GR even at $3^{rd}$ order post Newtonian approximation.

\begin{figure}
    \centering
     \includegraphics[width=0.65\columnwidth]{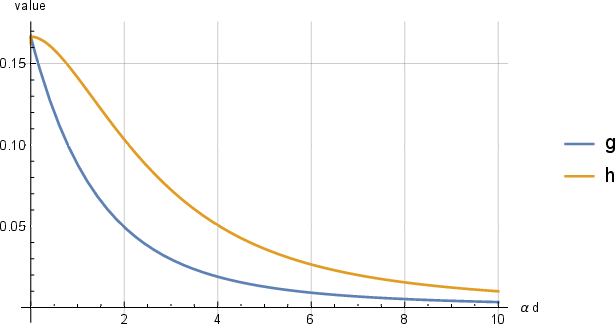}
    \caption{The evaluation of the functions $g(\alpha d)$ and $h(\alpha d)$ respect to $\alpha d$ }\label{fig:gh}
\end{figure}

However the difference $\Delta n$ will not vanish in a non-axially-symmetrical spacetime region, thus one could detect the difference in principle.
Since the potential appearing in the expression (\ref{eq:n_diff_axis}) without function $g$ are $3^{rd}$ post Newtonian terms, the order of $\Delta n$ is determined by the order of the functions $g(\alpha d)$ (or $g(\alpha d)$ and $h(\alpha d)$, with the same order as shown in Fig. \ref{fig:gh}. It is shown that, for $\alpha \leq {\frac{1}{d}}$, $\Delta n$ will be around $10^{-1}$ times the $3^{rd}$ GR terms, and thus keeps the same $3^{rd}$ order. However for $\alpha \geq \frac{10}{d}$, $\Delta n$ will be less than $10^{-2}$ times the $3^{rd}$ order terms, and hence is indistinguishable with the $4^{th}$ order term.
This estimation would approach the exact result for the spacetime region far away from the matter center. Then the functions $g$ and $h$ could be approximated by the values $g(\alpha d_0)$ and $h(\alpha d_0)$, where $d_0$ is the distance of the position to the matter center. Thus one could write the first-order approximation of (\ref{eq:n_diff_com}) as
\begin{equation}\label{eq:z_diff_app}
 \begin{split}
   \Delta n_j \simeq &  g(\alpha d_0)\frac{G }{c^5}\int_{c}{\frac{{^{(-2)}T^{00}(\bm{x}',t)} v'_j }{\vert \bm{x}-\bm{x}' \vert} } \td^3 x'-h(\alpha d_0)\frac{G }{c{^5}}\int_{c}{ \frac{{^{(-2)}T^{00}} v' \cdot (\bm{x}-\bm{x}')(\bm{x}-\bm{x}')_j}{\vert \bm{x}-\bm{x}' \vert ^3} }\td^3 x' \;,\\
 \end{split}
\end{equation}
in which the whole integration is approximated by the integration of the region around the center where most part of the matter locate.
Therefore, in highly non-axially symmetric spacetime region, it is possible to distinguish the lensing of GR from those of the $f(R)$ theories satisfying $\alpha \leq \frac{1}{d_0}$. It also requires that the measurement can approach the $10^{-1}$ precision of the $3^{rd}$ order effect.
\section{Concluding Remarks}\label{ch:summary}

In this paper, the post-Newtonian approximation of a general class of $f(R)$ theories is formulated up to $3^{rd}$ order. At the $3^{rd}$ order expansion, the metric components contain not only the Newtonian potential $U$ but also the Yukawa potential $V$ together with the third order potentials. Note that $f(R)$ theories can be transformed into generalized Brans-Dicke theories by suitable conformal transformations. Since the post-Newtonian formulation of Brans-Dicke gravity has been well studied \cite{will1993theory}, one can check the consistency of the post-Newtonian formulations between the two kinds of theories. It turns out that, in the limit of $\alpha \to 0$, our result (\ref{eq:fr_metric}) of $f(R)$ gravity coincides with the result of Brans-Dicke gravity given in \cite{will1993theory}. The proof will be presented in Appendix A.

 In our post-Newtonian formulation, the information of a specific form of $f(R)$ theories is contained in the Yukawa potential. While the Yukawa potential does not show in the $2^{nd}$ order expression of the effective refraction index $n$ of light, it does appear in the $3^{rd}$ order expression of $n$. Therefore in principle we could distinguish $f(R)$ gravity and GR. Moreover, detailed analysis shows that a series of $f(R)$ forms, more specific, whose parameter $1/\alpha \sim \sqrt{f''(0)/f'(0)}$ is larger than the distance to the massive center, are distinguishable from GR by the gravitational lensing effect at the $3^{rd}$ order post-Newtonian approximation. It should be noted that the conclusion that $f(R)$ theories can lead to the gravitational lensing effect different from that of GR can also be obtained by the approach of Minkowski functionals \cite{Chenxiaoji:2014mxa}. {However, it is shown in this paper that, in the axially symmetrical spacetime region, the gap term between these two kinds of theories vanishes and hence they are indistinguishable at $3^{rd}$ order.}
%cleared

One of the motivations for developing modified gravity theories is to account for the observed mass profiles in galaxies as well as clusters of galaxies without the inclusion of dark matter. The existence of dark matter in GR is confirmed by the observational data not only from the dynamical analysis, such as rotation curves in spiral galaxies \cite{Sofue:2000jx} and velocity dispersions in early-type systems \cite{Napolitano:2008ft,romanowsky2009mapping}, but also from gravitational lensing observations \cite{Auger:2009hj,Tortora:2010sy}. Observations indicate that we need to take into account almost the same large amount of dark matter to explain the gravitational lensing effect as that for the dynamical data like the velocity dispersion or the temperature profile of the X-ray emitting intracluster medium \cite{DeFilippis:2004vy,DeFilippis:2005hx} in galaxy clusters or spiral galaxies. Up to now, certain $f(R)$ theories are tested by the dynamical data in galaxy clusters and spiral galaxies \cite{nucamendi2001alternative, sobouti2007mathsf, Martins:2007uf,Capozziello:2008ny, capozziello2009testing,capozziello2012dark}. However, concerning the  gravitational lensing observations, our results here shows a disfavor of the attempts in this direction. For any $f(R)$ form which could be weakly expanded, the lensing effect correction due to the $f(R)$ from will be at most the $3^{rd}$ order, which is at most $10^{-2}$ times of the leading order, i.e., the $2^{nd}$ order post Newtonian effect. {{Moreover, the fact that in axially symmetrical spacetime region there is no difference in lensing effect between these two kinds of theories strongly indicates that most of the lensing observations will not show the difference even at $3^{rd}$ order. Thus it is impossible to explain the lensing observations in the pure $f(R)$ theories that we are considering without any dark matter involved.
%cleared

It is still possible to determine the parameter $\alpha^2:=f'(0)/(3f''(0))$ though the precise observational results in non-axially symmetric system. Thus in near future, precise observations of lensing effect would be useful  to distinguish certain $f(R)$ theories from GR. It should be remarked that our result is only valid for the $f(R)$ forms which could be weakly expanded. It is interesting to further study whether the dark matter content can be replaced by other unexpandable $f(R)$ theories or other kinds of modified gravity.
%cleared

\begin{acknowledgments}
This work is supported in part by the NSFC (grant nos.11235003 and 11475023), the Research Fund for the Doctoral Program of Higher Education of China, and the National Undergraduate Training Programs for Innovation of China.
\end{acknowledgments}

\appendix
\section{PPN formalism}
We will derive explicitly the $3^{rd}$ order terms and get its limit of the largest effect in this appendix. Especially, we could get the limit of $\alpha \to 0$ and compare it to the result with the  standard parametrized post Newtonian (PPN) form.

First we will write out explicitly $3^{rd}$ order $g_{ij}$ term, more precisely, the $Z_{0i}$ term containing in the $g_{ij}$ expression. The other two terms $Y_i$ and $\psi_{,0i}$ is the usual potential in post Newtonian form which reads
\begin{align}
Y_i(\bm{x},t):=&\frac{G}{c}\int{\frac{{^{(-1)}T^{0i}}(\bm{x}',t)}{\vert \bm{x}-\bm{x}' \vert}} \td^3x'=\frac{G}{c^2}\int{\frac{{^{(-2)}T^{00}(\bm{x}',t)} v'_j }{\vert \bm{x}-\bm{x}' \vert} \td^3x} \;,\\
\psi_{,0i}=&Y_i-W_i \;,
\end{align}
where $W_i$ reads
\begin{equation}
  W_i=\frac{G}{c{^2}}\int{ \frac{{^{(-2)}T^{00}(\bm{x}',t)} v' \cdot (\bm{x}-\bm{x}')(\bm{x}-\bm{x}')_j}{\vert \bm{x}-\bm{x}' \vert ^5} \td^3x} \;.
\end{equation}
From (\ref{eq:V}) and (\ref{eq:U}), we have
\begin{align}
U(\bm{x},t)_{,0j}:=& \frac{G}{c^2}\int{ \frac{{^{(-2)}T^{00}(\bm{x}',t)} v'_j}{\vert \bm{x}-\bm{x}' \vert ^3} } \td^3 x'-\frac{3 G}{c^2}\int{ \frac{{^{(-2)}T^{00}(\bm{x}',t)} v' \cdot (\bm{x}-\bm{x}')(\bm{x}-\bm{x}')_j}{\vert \bm{x}-\bm{x}' \vert ^5} } \td^3 x' \;,\\
\notag V(\bm{x},t)_{,0j}=&\frac{G}{c^2}\int{\frac{{^{(-2)}T^{00}(\bm{x}',t)}\mathrm{e}^{-\alpha \vert \bm{x}-\bm{x}' \vert}  v'_j (1 + \alpha \vert \bm{x}-\bm{x}' \vert)}{\vert \bm{x}-\bm{x}' \vert^3}} \td^3 x'\\
&-\frac{3G}{c^2}\int{\frac{{^{(-2)}T^{00}(\bm{x}',t)}\mathrm{e}^{-\alpha \vert \bm{x}-\bm{x}' \vert}  v' \cdot (\bm{x}-\bm{x}')(\bm{x}-\bm{x}')_j (1 + \alpha \vert \bm{x}-\bm{x}' \vert)}{\vert \bm{x}-\bm{x}' \vert^5}} \td^3 x'\\
\notag &-\frac{G}{c^2}\int{\frac{{^{(-2)}T^{00}(\bm{x}',t)}\mathrm{e}^{-\alpha \vert \bm{x}-\bm{x}' \vert} \alpha^2 v' \cdot (\bm{x}-\bm{x}') (\bm{x}-\bm{x}')_j}{\vert \bm{x}-\bm{x}' \vert^3}} \td^3 x' \;,
\end{align}
where $v'_j$ donates the $j$ components of the velocity at point $\bm{x}'$.

Thus one get
\begin{equation}\label{eq:z_diff_App}
\begin{split}
Z_{0j}:=&\frac{G}{c^2}\int {\frac{{^{(-2)}T^{00}(\bm{x}',t)} v'_j }{\vert \bm{x}-\bm{x}' \vert^3} } \frac{1}{3\alpha^2}(1-\mathrm{e}^{-\alpha \vert \bm{x}-\bm{x}' \vert} (1 + \alpha \vert \bm{x}-\bm{x}' \vert))\\
&+\frac{{^{(-2)}T^{00}(\bm{x}',t)} v' \cdot (\bm{x}-\bm{x}')(\bm{x}-\bm{x}')_j}{\alpha^2 \vert \bm{x}-\bm{x}' \vert ^5}\left(\mathrm{e}^{-\alpha \vert \bm{x}-\bm{x}'\vert} \bigg(1+\alpha \vert \bm{x}-\bm{x}' \vert +\frac{1}{3}\alpha^2 \vert \bm{x}-\bm{x}' \vert^2 \bigg)-1 \right) \td^3 x' \;.\\
\end{split}
\end{equation}

Then we are going to see the limit of $\alpha \to 0$. It is directly to see,
$\lim_{\alpha \to 0}V=U$.
Thus the limit of $g_{00}$ and $g_{ij}$ could be read out directly:
\begin{align}
    &g_{00}=-1+2\frac{U}{c^2} \frac{4}{3}, &g_{ij}&=(1+2\frac{U}{c^2} \frac{2}{3}) \delta_{ij} \;.
\end{align}
And one could show that, the limit of $\alpha \to 0$ leads to
\begin{equation}
  \begin{split}
    &\lim_{\alpha \to 0} \frac{1}{3\alpha^2}(1-\mathrm{e}^{-\alpha \vert \bm{x}-\bm{x}' \vert})=\frac{\vert \bm{x}-\bm{x}' \vert^2}{6} \;,\\
    &\lim_{\alpha \to 0} \frac{1}{\alpha^2}\left(\mathrm{e}^{-\alpha \vert \bm{x}-\bm{x}'\vert} \bigg(1+\alpha \vert \bm{x}-\bm{x}' \vert +\frac{1}{3}\alpha^2 \vert \bm{x}-\bm{x}' \vert^2 \bigg) \right)=-\frac{\vert \bm{x}-\bm{x}' \vert^2}{6} \;.
    \end{split}
\end{equation}
Thus for potential $Z_{0i}$ we simply get
\begin{equation}
\begin{split}
Z_{0j}:=&\frac{G}{c^2}\int{\frac{{^{(-2)}T^{00}(\bm{x}',t)} v'_j }{\vert \bm{x}-\bm{x}' \vert} }  \td^3 x'-\frac{G}{c{^2}}\int{ \frac{{^{(-2)}T^{00}(\bm{x}',t)} v' \cdot (\bm{x}-\bm{x}')(\bm{x}-\bm{x}')_j}{\vert \bm{x}-\bm{x}' \vert ^3} } \td^3 x'\\
=&\frac{1}{6}(Y_j-W_j) \;.
\end{split}
\end{equation}
Since from the standard PPN formalism in GR, we already know that
\begin{equation}
\begin{split}
g_{0i}^{GR}:=&\frac{1}{c^3}(-4Y_j +1/2 \psi_{,0i})=\frac{1}{c^3}(-\frac{7}{2}Y_j-\frac{1}{2}W_j) \;,
\end{split}
\end{equation}
we have
\begin{equation}
\begin{split}
g_{0i}:=&\frac{1}{c^3}(-\frac{10}{3}Y_j-\frac{2}{3}W_j) \;.
\end{split}
\end{equation}

After the redefinition of gravitational constant $G'=\frac{4}{3}G$ based on $g_{00}$, we get
\begin{equation}\label{eq:fr_metric PPN}
\begin{split}
{g_{00}}&=-1+\frac{2}{c^2}U \;,\\
{g_{0i}}&=\frac{1}{c^3}(-\frac{5}{2}Y_j-\frac{1}{2}W_j) \;,\\
{g_{ij}}&=(1+\frac{1}{c^2}U){\delta_{ij}} \;.
\end{split}
\end{equation}
and one directly read out $\gamma=0.5$ and $\alpha_1=\alpha_2=\zeta_1=\zeta=0$ respectively, which meet the result of this kind of $f(R)$ form given by C. M. Will \cite{will1993theory}.

%\bibliographystyle{apsrev}
 %\bibliography{refs}
 %

\end{document}